\documentstyle[preprint,aps]{revtex}

\draft
\newcommand{\be}{\begin{equation}}
\newcommand{\ee}{\end{equation}}
\newcommand{\ben}{\begin{eqnarray}}
\newcommand{\een}{\end{eqnarray}}

\newcommand{\iii}{\'{\i}}

\begin{document}

\draft
\title{Conditional $q$-Entropies and Quantum Separability:
A Numerical Exploration}
\author{J. Batle$^1$,  A. R. Plastino$^{1,\,2}$, M. Casas$^1$,
and A. Plastino$^{2}$}

\address {$^1$Departament de F\iii sica, Universitat de les Illes Balears,
07071 Palma de Mallorca, Spain \\ $^2$National University La Plata
and  Argentina's National Research Council (CONICET), C.C. 727,
1900 La Plata, Argentina}


\maketitle
 \begin{abstract}

 We revisit the relationship between quantum separability and the sign
 of the relative $q$-entropies of composite quantum systems. The
 $q$-entropies depend on the density matrix eigenvalues $p_i$ through
 the quantity $\omega_q = \sum_i p_i^q$. R\'enyi's and Tsallis' measures
 constitute particular instances of these entropies. We perform a
 systematic numerical survey of the space of mixed states of two-qubit
 systems in order to determine, as a function of the degree of
 mixture, and for different values of the entropic parameter $q$,
 the volume in state space occupied by those states characterized by
 positive values of the relative entropy. Similar calculations are performed
 for qubit-qutrit systems and for composite systems described by Hilbert
 spaces of larger dimensionality. We pay particular attention to the
 limit case $q\rightarrow \infty$. Our numerical results indicate
 that, as the dimensionalities of both subsystems increase,
 composite quantum systems tend, as far as their relative
 $q$-entropies are concerned, to behave in a classical way.

\noindent
 Pacs: 03.67.-a; 89.70.+c; 03.65.Bz


\end{abstract}

\maketitle

\newpage

\section{Introduction}

 Important steps have been recently made towards a systematic
exploration of the space of arbitrary (pure or mixed) states of composite
quantum systems \cite{ZHS98,Z99,Z01} in order to determine the typical features
exhibited by these states with regards to the phenomenon of quantum
entanglement \cite{ZHS98,Z99,Z01,MJWK01,IH00,BCPP02a,BCPP02b}. This phenomenon
is one of the most fundamental and non-classical features exhibited by quantum
systems \cite{LPS98}. Quantum entanglement lies at the basis of some of the
most important processes studied by quantum information theory
\cite{LPS98,WC97,W98,NC00,GD02}, such as quantum cryptographic key distribution
\cite{E91}, quantum teleportation \cite{BBCJPW93}, superdense coding
\cite{BW93}, and quantum computation \cite{EJ96,BDMT98}. A state of a composite
quantum system is called ``entangled" if it can not be represented as a mixture
of factorizable pure states. Otherwise, the state is called separable. The
above definition is physically meaningful because entangled states (unlike
separable states) cannot be prepared locally by acting on each subsystem
individually \cite{P93}.

  When one deals with a classical composite system, described by a
 suitable probability distribution defined over the concomitant
 phase space, the entropy of any of its subsystems is always equal
 or smaller than the entropy characterizing the whole system. This
 is also the case for separable states of a composite quantum
 system \cite{NK01,VW02}. In contrast,
 a subsystem of a quantum system described by an
 entangled state may have an entropy greater than the entropy of
 the whole system. In point of fact, the von Neumann entropy of either
 of the subsystems of a bipartite quantum system described (as a whole)
 by a pure state provides a natural measure of the amount of entanglement
 of such state. Thus, a pure state (which has vanishing entropy)
 is entangled if and only if its subsystems have an entropy
 larger than the one associated with the system as a whole.
  The situation is more complex when the composite system is
  described by a mixed state. As already mentioned, there are
  entangled  mixed states such that the entropy of the complete
  system is smaller than the entropy of one of its subsystems.
  Alas, entangled mixed states such that the entropy of the
  system as a whole is larger than the entropy of either of
  its subsystems exist as well. Consequently, the classical
  inequalities relating the entropy of the whole system with the
  entropies of its subsystems provide only necessary, but not
  sufficient, conditions for quantum separability.
  There are several entropic (or information) measures that can be
  used in order to implement these entropic criteria for separability.
  Considerable attention has been paid, in this regard, to the
  $q$-entropies
  \cite{VW02,HHH96,HH96,CA97,V99,TLB01,TLP01,T02,A02},
  which incorporate
  both R\'enyi's \cite{BS93} and Tsallis' \cite{T88,LV98,LSP01} families of
  information measures as special instances (both admitting, in turn,
  Shanon's measure as the particular case associated with the limit
  $q\rightarrow 1$). The early motivation for these studies was
  the development of practical separability criteria for density matrices.
  The discovery by Peres of the partial transpose criteria, which for
  two-qubits and qubit-qutrit systems turned out to be both necessary
  and sufficient, rendered that original motivation somewhat outmoded.
  In point of fact, it is not possible to find a necessary and sufficient
   criterium for separability based
  solely upon the eigenvalue spectra of the three density matrices
  $\rho_{AB}, \rho_A=Tr_B[\rho_{AB}]$, and $\rho_B=Tr_A[\rho_{AB}]$
  associated with a composite system $A\oplus B$ \cite{NK01}.
  However, {\it the violation of classical entropic inequalities by
  entangled quantum states is of considerable interest in its own right}.
  Quantum entanglement is a fundamental aspect of quantum physics
  that deserves to be investigated in full detail from all possible
  points of view. The violation of the classical entropic
  inequalities provides a clear and direct information-theoretical
  manifestation of the phenomenon of entanglement.

  The aim of the present work is to study the relationship between
 quantum separability and the violation of the classical $q$-entropic
 inequalities (which corresponds to  negative values of the relative
 $q$-entropies). We will perform a systematic numerical survey of the space
 of mixed states of two-qubit systems in order to determine, as a
 function of the degree of mixture, and for different values of the
 entropic parameter $q$, the volume in state space occupied by those
 states characterized by positive values of the relative $q$-entropies.
 Similar calculations are performed for qubit-qutrit systems and for
 composite systems described by Hilbert spaces of larger dimensionality.
 We pay particular attention to the limit case $q\rightarrow \infty$.

  The paper is organized as follows. In section II we review some basic
 properties of the $q$-entropies and the relative $q$-entropies. Our main
 results are discussed in sections III. Finally, some conclusions are
 drawn in section IV.

   \section{q-Entropies and q-Relative Entropies}

  There are several entropic (or information) measures that can
  be useful in order to investigate the violation of classical entropic
  inequalities by quantum entangled states. The von Neumann measure

  \be \label{slog}
  S_1 \, = \,- \, Tr \left( \hat \rho \ln \hat \rho \right),
  \ee

  \noindent
  is important because of its relationship with the thermodynamic
  entropy. On the other hand, the so called participation
  ratio,

  \be \label{partrad}
  R(\hat \rho) \, = \, \frac{1}{Tr(\hat \rho^2)},
  \ee

  \noindent
  is particularly convenient for calculations \cite{ZHS98,MJWK01,ZZF00}.
  The $q$-entropies, which are functions of the quantity

  \be \label{trq}
  \omega_q \, = \, Tr \left( \hat \rho^q \right),
  \ee

  \noindent
  provide one with a whole family of entropic measures.
  In the limit $q\rightarrow 1 $ these measures incorporate (\ref{slog})
  as a particular instance. On the other hand, when $q=2$ they are
  simply related to the participation ratio (\ref{partrad}). Most of the
  applications of $q$-entropies to physics involve either the R\'enyi entropies
  \cite{BS93},

  \be \label{renyi}
   S^{(R)}_q \, = \, \frac{1}{1-q} \, \ln \left( \omega_q \right),
  \ee

  \noindent
  or the Tsallis' entropies \cite{T88,LV98,LSP01}

  \be \label{tsallis}
  S^{(T)}_q \, = \, \frac{1}{q-1}\bigl(1-\omega_q \bigr).
  \ee

  \noindent We reiterate that the von Neumann measure (\ref{slog}) constitutes a particular
  instance of both R\'enyi's and Tsallis' entropies, which is obtained in the
  limit $q\rightarrow 1$. The most distinctive single property of
  Tsallis' entropy is its nonextensivity. The Tsallis' entropy of a
  composite system $A \oplus B $ whose state is described
  by a factorizable density matrix, $\rho_{AB} = \rho_A \otimes
  \rho_B$, is given by Tsallis' $q$-additivity law,

  \be
  S_q^{(T)}(\rho_{AB}) \, = \, S_q^{(T)}(\rho_A) \, + \, S_q^{(T)}(\rho_B)
  \, + \, (1-q)S_q^{(T)}(\rho_A)S_q^{(T)}(\rho_B).
   \ee

  \noindent
   In contrast, R\'enyi's entropy is extensive. That is,
   if $\rho_{AB} = \rho_A \otimes  \rho_B$,

  \be
  S_q^{(R)}(\rho_{AB}) \, = \, S_q^{(R)}(\rho_A) \, + \,
  S_q^{(R)}(\rho_B).
  \ee

  \noindent
  Tsallis' and R\'enyi's measures are
  related through

  \be \label{stsr}
  S^{(T)}_q \, = \,F( S^{(R)}_q),
  \ee

  \noindent
  where the function $F$ is given by

  \be \label{fx}
  F(x) \, = \, \frac{1}{1-q} \left\{ e^{(1-q)x} - 1 \right\}.
  \ee

  \noindent
  An immediate consequence of equations (\ref{stsr}-\ref{fx})
  is that, for all non vanishing values of $q$, Tsallis' measure
  $ S^{(T)}_q$ is a monotonic increasing function of R\'enyi's
  measure $ S^{(R)}_q $.

  Considerably attention has been recently paid to a relative
  entropic measure based upon Tsallis' functional, and defined as

  \be \label{qurela}
  S^{(T)}_q(A|B) \, = \,
  \frac{S^{(T)}_q(\rho_{AB})-S^{(T)}_q(\rho_B)}{1+(1-q)S^{(T)}_q(\rho_B)}.
  \ee

  \noindent
  Here $\rho_{AB}$ designs an arbitrary quantum state of the
  composite system $A\oplus B$,
  not necessarily factorizable nor separable,
  and $\rho_B = Tr_A (\rho_{AB})$. The relative $q$-entropy
  $S^{(T)}_q(B|A)$ is defined in a similar way as (\ref{qurela}),
  replacing $\rho_B $ by $\rho_A = Tr_B (\rho_{AB})$.
     The relative $q$-entropy
  (\ref{qurela}) has been recently studied in connection
  with the separability of density matrices describing composite
  quantum systems \cite{TLB01,TLP01}. For separable states,
  we have \cite{VW02}

  \ben \label{qsepar}
  S^{(T)}_q(A|B) &\ge & 0, \cr
  S^{(T)}_q(B|A) &\ge & 0.
  \een

  \noindent
  On the contrary, there are entangled states that have negative
  relative $q$-entropies. That is, for some entangled states one (or both)
  of the inequalities (\ref{qsepar}) are not verified.

  Notice that the denominator in (\ref{qurela}),

  \be \label{deno}
  1+(1-q)S^{(T)}_q \, = \, w_q \, > \, 0.
  \ee

  \noindent
  is always positive. Consequently, as far as the sign of the
  relative entropy is concerned, the denominator in (\ref{qurela})
  can be ignored. Besides, since Tsallis' entropy is a monotonous
  increasing function of R\'enyi's (see Equations (\ref{stsr}-\ref{fx})),
  it is plain that (\ref{qurela}) has always the same sign as

  \be \label{relarenyi}
  S^{(R)}_q(A|B) \, = \, S^{(R)}_q(\rho_{AB})-S^{(R)}_q(\rho_{B}).
  \ee

  \noindent
  From now on we are going to refer to the positivity of either
 the Tsallis' relative entropy (\ref{qurela}) or the R\'enyi
 relative entropy (\ref{relarenyi}) as the ``classical $q$-entropic
 inequalities". In general, when we  speak about the sign of the
 $q$-relative entropy, we are going to refer indistinctly either to
 the sign of (\ref{qurela}) or to the sign of (\ref{relarenyi})
 (which always coincide).

\section{Probabilities of finding states with positive relative $q$-entropies.}


In order to perform a systematic numerical survey of the
properties of arbitrary (pure and mixed) states of a given quantum
system, it is necessary to introduce an appropriate measure $\mu $
on the concomitant space ${\cal S}$ of general quantum states.
Such a measure is needed to compute volumes within the space
${\cal S}$, as well as to determine what is to be understood by a
uniform distribution of states on ${\cal S}$.  A natural measure
on ${\cal S}$, which we are going to adopt in the present work,
was recently introduced by Zyczkowski {\it et al.}
\cite{ZHS98,Z99}.
 An arbitrary (pure or mixed) state $\rho$ of a quantum system
described by an $N$-dimensional Hilbert space can always be
expressed as the product of three matrices,

\be \label{udot} \rho \, = \, U D[\{\lambda_i\}] U^{\dagger}. \ee

\noindent Here $U$ is an $N\times N$ unitary matrix and
$D[\{\lambda_i\}]$ is an $N\times N$ diagonal matrix whose
diagonal elements are $\{\lambda_1, \ldots, \lambda_N \}$, with $0
\le \lambda_i \le 1$, and $\sum_i \lambda_i = 1$.
  The group of unitary matrices $U(N)$ is
endowed with a unique, uniform measure: the Haar measure $\nu$
\cite{PZK98}. On the other hand, the $N$-simplex $\Delta$,
consisting of all the real $N$-uples $\{\lambda_1, \ldots,
\lambda_N \}$ appearing in (\ref{udot}), is a subset of a
$(N-1)$-dimensional hyperplane of ${\cal R}^N$. Consequently, the
standard normalized Lebesgue measure ${\cal L}_{N-1}$ on ${\cal
R}^{N-1}$ provides a natural measure for $\Delta$. The
aforementioned measures on $U(N)$ and $\Delta$ lead then to a
natural measure $\mu $ on the set ${\cal S}$ of all the states of
our quantum system \cite{ZHS98,Z99,PZK98}, namely,

\be \label{memu}
 \mu = \nu \times {\cal L}_{N-1}.
 \ee

 \noindent
  All our present considerations are based on the assumption
 that the uniform distribution of states of a quantum system
 is the one determined by the measure (\ref{memu}). Thus, in our
 numerical computations we are going to randomly generate
 states according to the measure (\ref{memu}).

    The simplest quantum mechanical systems exhibiting the phenomenon of
 entanglement are two-qubits systems ($N=4$). They play a fundamental role in
 Quantum Information Theory. The concomitant space of mixed states is
 $15$-dimensional and its properties are not trivial. There still are features
 of this state space, related to the phenomenon of entanglement, which have
 not, thus far, been completely characterized in full detail.

 We determined numerically, by recourse to a Monte Carlo calculation
 and for different values of the entropic parameter $q$,
 the probability of finding a two-qubits state which,
 for a given degree of mixture $R= 1/Tr \, (\rho^2)$,
 has positive relative $q$-entropies
 (i.e., $S^{(R)}_q(\rho_{AB})\ge S^{(R)}_q(\rho_{A})$ and
 $S^{(R)}_q(\rho_{AB})\ge S^{(R)}_q(\rho_{B})$). The results
are depicted in Fig. 1. The curve associated with the limit case
$q\rightarrow \infty$ deserves special comment. In this limit we
have,

\be \label{limiqinf} \lim_{q\to \infty } \, \left( Tr \rho^q
\right)^{1/q} \, = \,\lim_{q\to \infty } \, \left( \sum_i
p_i^q\right)^{1/q} \, = \, \lambda_m, \ee

\noindent where

\be \lambda_m = \max_{i} \{ p_i \} \ee

\noindent is the maximum eigenvalue of the statistical operator
$\rho$. Hence, in the limit $q\to \infty $, the $q$-entropies
depend only on the largest eigenvalue of the density matrix. In
particular, the R\'enyi entropy reduces to

\be \label{reninf} S^{(R)}_{\infty} \, = \, -\ln \left( \lambda_m
\right).
 \ee

\noindent This means that the curve in Fig. 1 associated with
$q=\infty$ indicates the probabilities of finding states such that
the largest eigenvalue of the statistical operator describing the
composite system is smaller than the largest eigenvalues of either
of its subsystems. The solid line in Fig. 1 corresponds to the
probability of finding, for a given degree of mixture $R= 1/Tr \,
(\rho^2)$, a two-qubits state with a positive partial transpose.
Since Peres' criterium for separability is necessary and
sufficient, this last probability coincides with the probability
of finding a separable state. We see that, as the value of $q$
increases, the curves associated with the relative entropies
approaches the curve corresponding to Peres criterium. However,
even in the limit $q \rightarrow \infty$ the entropic curve lies
above the Peres' one by a considerable amount. This means that,
even for $q \rightarrow \infty$, {\it there is a considerable
volume in state space occupied by entangled states complying with
the classical entropic inequalities} (that is, having positive
relative entropies).

   The probability of finding separable states increases with the
degree of mixture \cite{ZHS98}, as it is evident from the solid
curve in Fig. 1. Also, one can appreciate the fact that a similar
trend is exhibited by the probability of finding, for a given
$q$-value, states with positive relative $q$-entropies.

 We have computed numerically the probability (for different
 values of $q$) that a two-qubits state with a given degree of
 mixture be correctly classified, either as entangled or as separable,
 on the basis of the sign of the relative $q$-entropies. The results
 are plotted in Fig. 2. That is, Fig. 2 depicts the probability of
 finding (for different values of $q$) a two-qubits state which,
 for a given degree of mixture $R= 1/Tr \, (\rho^2)$, either has
 (i) both relative $q$-entropies positive, as well as a positive partial transpose,
 or (ii) has a negative relative $q$-entropy and a non positive partial
 transpose. We see that, for all values of $q>0$,
  this probability is equal to one both for pure
 states ($R=1$) and for states with ($R>3$). The probability
 attains its lowest value $P_m(q)$ at a special value $R_m(q)$ of the
 participation ratio. Both quantities $R_m(q)$ and $P_m(q)$
 exhibit a monotonic increasing behaviour with $q$, adopting their
 maximum values in the limit $q\rightarrow \infty$.

In Fig. 1 and Fig. 2 we have used the participation ratio $R$ as a
measure of mixedness. The quantity $R$ is, essentially, a
$q$-entropy with $q=2$. The $q$-entropies associated with other
values of $q$ are legitimate measures of mixedness as well, and
have already found applications in relation with the study of
entanglement \cite{ZHS98,BCPP02b}. It is interesting to see what
happens, in the present context, when we consider measures of
mixedness based on other values of $q$. Of particular interests is
the limit case $q\rightarrow \infty$ which, as already mentioned,
is related to the largest eigenvalue of the density matrix. The
largest eigenvalue constitutes a legitimate measure of mixture in
its own right: states with larger values of $\lambda_m$ can be
regarded as less mixed. Its extreme values correspond to (i) pure
states (with $\lambda_m =1$) and (ii) the density matrix
$\frac{1}{4}\hat I$ (with $\lambda_m = 1/4$). In Figures 3 and 4
we have considered (in the horizontal axes) the largest eigenvalue
$\lambda_m$ as a measure of mixedness.
 We computed the probability of finding (for different values
of $q$) a two-qubits state which, for a given value of the maximum
eigenvalue $\lambda_m$, has positive relative $q$-entropies. The
results are depicted in Fig. 3. The solid line corresponds to the
probability of finding, for a given degree of mixture $R= 1/Tr \,
(\rho^2)$, a two-qubits state with a positive partial transpose.
We see in Fig. 3 that, for $\lambda<1/3$, the probability of
finding states verifying the classical entropic inequalities
(i.e., having positive relative entropies) is, for all $q>0$,
equal to one. This is so because all states whose largest
eigenvalue $\lambda_m$ is less or equal than $1/3$ are separable
\cite{BCPP02b}.

Fig. 4 depicts the probability of finding (for different values of
$q$) a two-qubits state which, for a given value of the maximum
eigenvalue $\lambda_m$, either has (i) both relative $q$-entropies
positive and a positive partial transpose, or (ii) a negative
relative $q$-entropy and a non positive partial transpose.

  A remarkable aspect of the behaviour of the sign of the
relative $q$-entropies, which transpires from Figures 1 and 3, is
that, for any degree of mixture, the volume corresponding to
states with positive relative $q$-entropies ($q>0$) is a
monotonous decreasing function of $q$. This feature of Figures 1
and 3 is interesting because, for a single given state $\rho $,
the relative $q$-entropy is not necessarily decreasing in $q$
\cite{VW02}. This means that the positivity of the relative
entropy of a given state $\rho $ and for a given value $q^{*}$ of
the entropic parameter does not imply the positivity of the
relative $q$-entropies of that state for all $q<q{*}$. That is,
$q<q^{*}$ does not imply that the family of states exhibiting
positive relative $q^{*}$-entropies is a subset of the family of
states with positive $q$-entropies. This fact notwithstanding, the
numerical results reported here indicate that for $0<q<q{*}$ the
volume of states with positive $q^{*}$-relative entropies is
smaller than the volume of states with positive $q$-entropies.
This implies that, among all the $q$-entropic separability
criteria, the one corresponding to the limit case $q\rightarrow
\infty$ is the strongest one, as was recently suggested by Abe
\cite{A02} on the basis of his analysis of a monoparametric family
of mixed states for multi-qudit systems.

It is interesting to see the behaviour, as a function of the
entropic parameter $q$, of the global probability (regardless of
the degree of mixture) that an arbitrary state of a two-qubit
system exhibits simultaneously (i) a positive relative $q$-entropy
and a positive partial transpose, or (ii) a negative relative
$q$-entropy and a non positive partial transpose. In order words,
this is the probability that for an arbitrary state the entropic
separability criterium and the Peres' criterium lead to the same
``conclusion" with respect to the separability (or not) of the
state under consideration. In Fig. 5 we depict this probability as
a function of $1/q$, for values of $q\in [2,20]$. We see that this
probability is an increasing function of $q$. In the limit
$q\rightarrow \infty $ this probability approaches the value
$\approx 0.7428$. On the other hand, for $q=1$ (that is, when we
use the standard logarithmic entropy) the probability is
approximately equal to $0.6428$.

We have performed for qubit-qutrit systems calculations similar to
the ones that we have already discussed for two-qubits systems.
The results are summarized in Figures 6 and 7.  Fig. 6 depicts the
probability of finding (for different values of $q$) a
qubit-qutrit state which, for a given degree of mixture $R= 1/Tr
\, (\rho^2)$, has positive relative $q$-entropies. The solid line
in Fig. 6 corresponds to the probability of finding, for a given
degree of mixture $R= 1/Tr \, (\rho^2)$, a qubit-qutrit state with
a positive partial transpose. Fig. 7 depicts the probability of
finding, for different values of $q$, a qubit-qutrit state which
has, for a given degree of mixture $R= 1/Tr \, (\rho^2)$, either
(i) its two relative $q$-entropies positive, as well as a positive
partial transpose, or (ii)  a negative relative $q$-entropy and a
non positive partial transpose.  We have also computed the
probability (for different  values of $q$) that an arbitrary
qubit-qutrit state (regardless of its degree of mixture) be
correctly classified, either as entangled or as separable,
 on the basis of the sign of the relative $q$-entropies. These
 probabilities are depicted in Fig. 8, for values of $q$ in the
 interval $q\in [2,20]$. As happens with two-qubits systems, this
 probability is an increasing function of $q$. For $q=1$ the
 probability is approximately equal to $0.3891$ and approaches
 the (approximate) value $0.4974$ as $q\rightarrow \infty$. For
 a given value of $q$, the probability of coincidence between the
 Peres' and the entropic separability criteria are seen to be
 smaller in the case of qubit-qutrit systems than in the case of
 two-qubits systems.

   It is worth to investigate the manner in which the (negative)
relative $q$-entropy  $ S^{(R)}_q(\rho_{A})  \, - \,
S^{(R)}_q(\rho_{AB})$ is related to the entanglement of formation
\cite{BDSW96}, for general two-qubits states violating the
concomitant classical entropic inequality. We have studied the
aforementioned relationship numerically. The entanglement of
formation  of a two-qubits state $\hat \rho$ can be evaluated
analytically by recourse to Wootters' formula \cite{WO98},

\be
E[\hat \rho] \, = \, h\left( \frac{1+\sqrt{1-C^2}}{2}\right), \ee

\noindent where

\be
h(x) \, = \, -x \log_2 x \, - \, (1-x)\log_2(1-x), \ee

\noindent and the concurrence $C$ is given by

\be
C \, = \, max(0,\lambda_1-\lambda_2-\lambda_3-\lambda_4), \ee

\noindent $\lambda_i, \,\,\, (i=1, \ldots 4)$ being the square
roots, in decreasing order, of the eigenvalues of the matrix $\hat
\rho \tilde \rho$, with

\be \label{rhotil} \tilde \rho \, = \, (\sigma_y \otimes \sigma_y)
\rho^{*} (\sigma_y \otimes \sigma_y). \ee

\noindent The above expression has to be evaluated by recourse to
the matrix elements of $\hat \rho$ computed  with respect to the
product basis. In Fig. 9, the concurrence squared
 $C^2$ is plotted versus
 $ S^{(R)}_q(\rho_{A})  \, - \, S^{(R)}_q(\rho_{AB}),\,\,\, (q=\infty)$,
 for a set of random two-qubits states
 generated numerically, and keeping only those with a
 negative relative entropy. It can be appreciated in Fig. 9
 that, for those states not complying with the classical
 inequality $S^{(R)}_q(A|B) \ge  0$,
 the concurrence squared $C^2$ (and consequently,
 the entanglement of formation) is, to a certain extent,
 correlated with the relative $q$-entropy $ S^{(R)}_q(A|B)$.

 Finally, we have computed the probabilities of finding states
 with positive relative $q$-entropies (for the case
 $q=\infty$) for bipartite quantum systems described by Hilbert
 spaces of increasing dimensionality. Let $N_1$ and $N_2$ stand for
 the dimensions of the Hilbert spaces associated with each subsystem,
 and $N=N_1\times N_2 $ be the dimension of the Hilbert space
associated with the concomitant composite system. We have
considered three sets of systems: (i) systems with $N_1=2,3$ and
increasing values of $N_2$, and (ii) systems with $N_1=N_2$ and
increasing dimensionality. The computed probabilities  are
depicted in Figure 10, as a function of the total dimension $N$.
The three upper curves correspond (as indicated in the figure) to
composite systems with $N_1=2$, $N_1=3$, and $N_1=N_2$. For the
sake of comparison, the probability of finding  states complying
with the Peres partial transpose separability criterium (lower
curve) is also plotted. In order to obtain each point in Figure
10, $10^8$ states were randomly generated.

 Some interesting conclusions can be drawn from Figure 10.
 In the case of composite systems with $N_1=N_2$, the probability
 of finding states complying with the classical ($q=\infty $)
 entropic inequalities (that is, having positive both relative
 q-entropies) is an increasing function of the dimensionality.
 Furthermore, this probability seems to approach $1$, as
 $N\rightarrow \infty$. In other words, Figure 10 provides
 numerical evidence that, in the limit of infinite dimension,
 two-qudits systems behave classically, as far as the signs of the
 relative $q$-entropies are concerned.

 When considering composite systems with increasing dimensionality,
 but keeping the dimension of one of the subsystem constant
 ($N_1=2,3$), we obtained numerical evidence  that the probability
 of having positive relative $q$-entropies (again, with $q=\infty $)
 behave in a  monotonous decreasing way with the total dimension $N$.

  It is interesting to notice that the probabilities of finding
  states with positive $q$-entropies are not just a function
  of the total dimension $N=N_1\times N_2$ (as happens,
  with good approximation, for the probability of having a
  positive partial transpose). On the contrary, they
  depend on the individual dimensions ($N_1$ and $N_2$) of both
  subsystems. Furthermore, the trend of the alluded to
  probabilities are clearly different if one considers composite
  systems of increasing dimension with either (i) increasing
  dimensions for both subsystems or (ii) increasing dimension
  for one of the subsystems and constant dimension for the other
  one.

\section{Conclusions}

We have performed a systematic numerical survey of the space of
mixed states of two-qubit systems in order to determine, as a
function of the degree of mixture, and for different values of the
entropic parameter $q$, the volume in state space occupied by
those states characterized by positive values of the relative
$q$-entropy. We also computed, for different values of $q$, the
global probability of classifying correctly an arbitrary state of
a two-qubits system (either as separable or as entangled) on the
basis of the signs of its relative $q$-entropies. This probability
exhibits a monotonous increasing behaviour with the entropic
parameter $q$. The approximate values of these probabilities are
$0.6428$ for $q=1$ and $0.7428$ in the limit $q\rightarrow
\infty$.

An interesting conclusion that can be drawn from the numerical
results reported here is that, notwithstanding the known non
monotonicity in $q$ of the relative $q$-entropies \cite{VW02}, the
volume corresponding to states with positive relative
$q$-entropies ($q>0$) is, for any degree of mixture, a monotonous
decreasing function of $q$.

Similar calculations were performed for qubit-qutrit systems and
for composite systems described by Hilbert spaces of larger
dimensionality. We pay particular attention to the limit case
$q\rightarrow \infty$. Our numerical results indicate that, for
composite systems consisting of two subsystems characterized by
Hilbert spaces of equal dimension $N_1$, the probability of
finding states with positive $q$-entropies tend to 1 as $N_1$
increases. In oder words, as $N_1\rightarrow \infty $ most states
seem to behave (as far as their relative $q$-entropies are
concerned) classically.

 \acknowledgments This work was partially supported by
the  DGES grants PB98-0124  (Spain), and by CONICET (Argentine
Agency).

\newpage

\noindent {\bf FIGURE CAPTIONS}

 \vskip 0.5cm

\noindent Fig. 1- Probability of finding (for different values of
$q$) a two-qubits state which, for a given degree of mixture $R=
1/Tr \, (\rho^2)$, has positive relative $q$-entropies. The solid
line corresponds to the probability of finding, for a given degree
of mixture $R= 1/Tr \, (\rho^2)$, a two-qubits state with a
positive partial transpose.

\vskip 0.5cm

\noindent Fig. 2- Probability of finding (for different values of $q$) a
two-qubits state which, for a given degree of mixture $R= 1/Tr \, (\rho^2)$,
either (i) has both relative $q$-entropies positive, as well as a positive
partial transpose, or (ii) has a negative relative $q$-entropy and a non
positive partial transpose.

\vskip 0.5cm

\noindent Fig. 3- Probability of finding (for different values of
$q$) a two-qubits state which, for a given value of the maximum
eigenvalue $\lambda_m$, has positive relative $q$-entropies. The
solid line corresponds to the probability of finding, for a given
value of $\lambda_m$, a two-qubits state with a positive partial
transpose.

\vskip 0.5cm

\noindent Fig. 4- Probability of finding (for different values of
$q$) a two-qubits state which, for a given value of the maximum
eigenvalue $\lambda_m$, either (i) has its two relative
$q$-entropies positive, as well as a positive partial transpose,
or (ii) has a negative relative $q$-entropy and a non positive
partial transpose.

\vskip 0.5cm

\noindent Fig. 5- Probability (as a function of $q$) of finding a
two-qubits state which either has both  positive relative
$q$-entropies and a positive partial transpose, or has a negative
relative $q$-entropy and a non positive partial transpose.

\vskip 0.5cm

\noindent  Fig. 6- Probability of finding (for different values of
$q$) a qubit-qutrit state which, for a given degree of mixture $R=
1/Tr \, (\rho^2)$, has positive relative $q$-entropies. The solid
line corresponds to the probability of finding, for a given degree
of mixture $R= 1/Tr \, (\rho^2)$, a qubit-qutrit state with a
positive partial transpose.

\vskip 0.5cm

\noindent Fig. 7-  Probability of finding a qubit-qutrit state
which, for a given degree of mixture $R= 1/Tr \, (\rho^2)$, and
for different values of $q$, either (i) has its two relative
$q$-entropies positive, as well as a positive partial transpose,
or (ii) has a negative relative $q$-entropy and a non positive
partial transpose.

\vskip 0.5cm

\noindent Fig. 8- Probability (as a function of $q$) of finding a
qubit-qutrit state which either has both positive relative
$q$-entropies and a positive partial transpose, or has a negative
relative $q$-entropy and a non positive partial transpose.

\vskip 0.5cm

 \noindent Fig. 9- The concurrence squared
 $C^2$ is plotted versus
 $ S^{(R)}_q(\rho_{A})  \, -
 \, S^{(R)}_q(\rho_{AB}),\,\,\, (q=\infty)$,
 for a set of random two-qubits states
 generated numerically, keeping only those with a
 negative relative entropy.

 \vskip 0.5cm

\noindent  Fig. 10- Global probability of finding a state (pure or
mixed) of a bipartite quantum system with positive relative
$q$-entropies. $N_1$ and $N_2$ stand for the dimensions of the
Hilbert spaces associated with each subsystem, and $N=N_1\times
N_2 $ is the dimension of the Hilbert space associated with the
composite system as a whole. The three upper curves correspond (as
indicated in the figure) to composite systems of increasing
dimensionality, and with $N_1=2$, $N_1=3$, and $N_1=N_2$. The
probability of finding a state complying with the Peres partial
transpose separability ctiterium (lower curve) is also plotted.

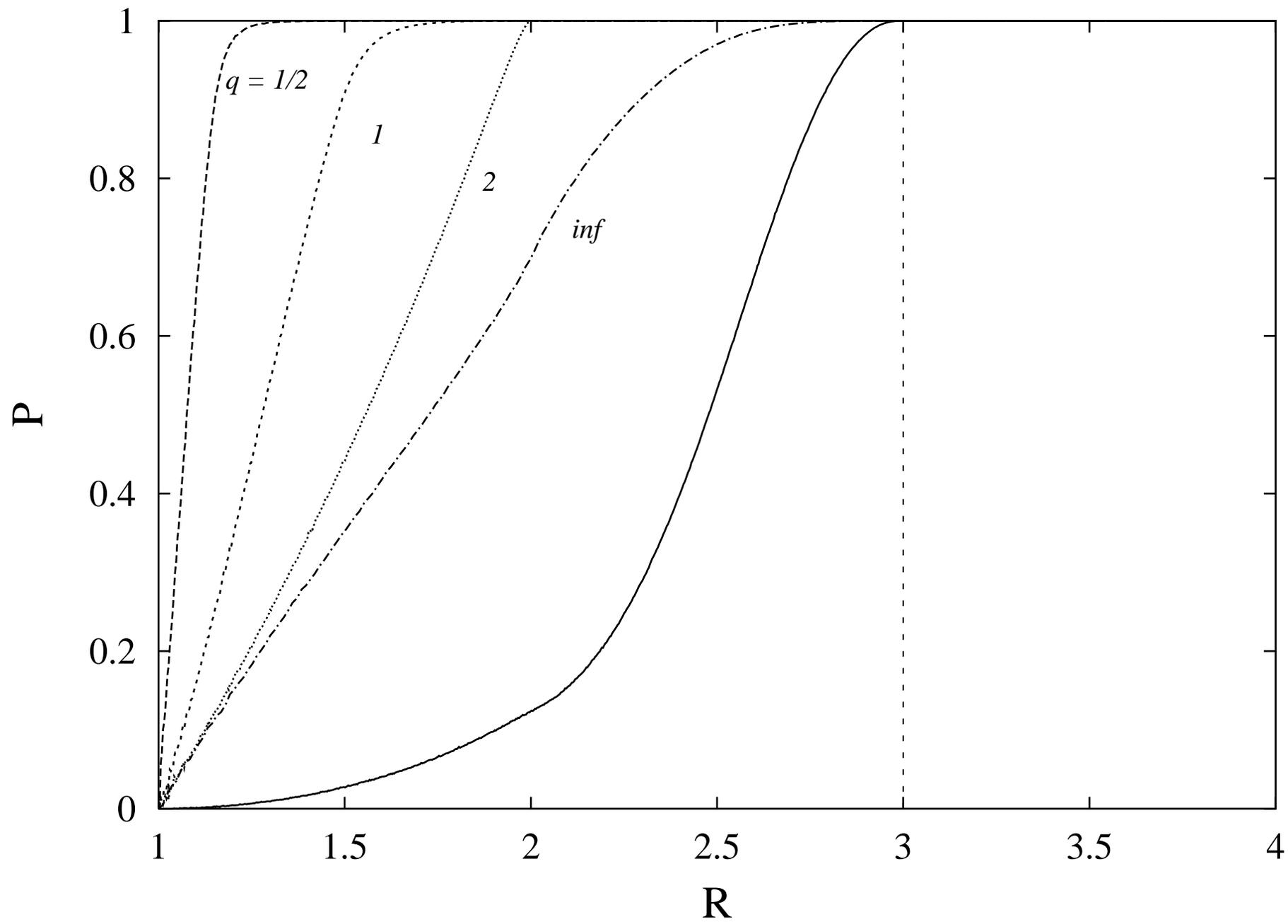
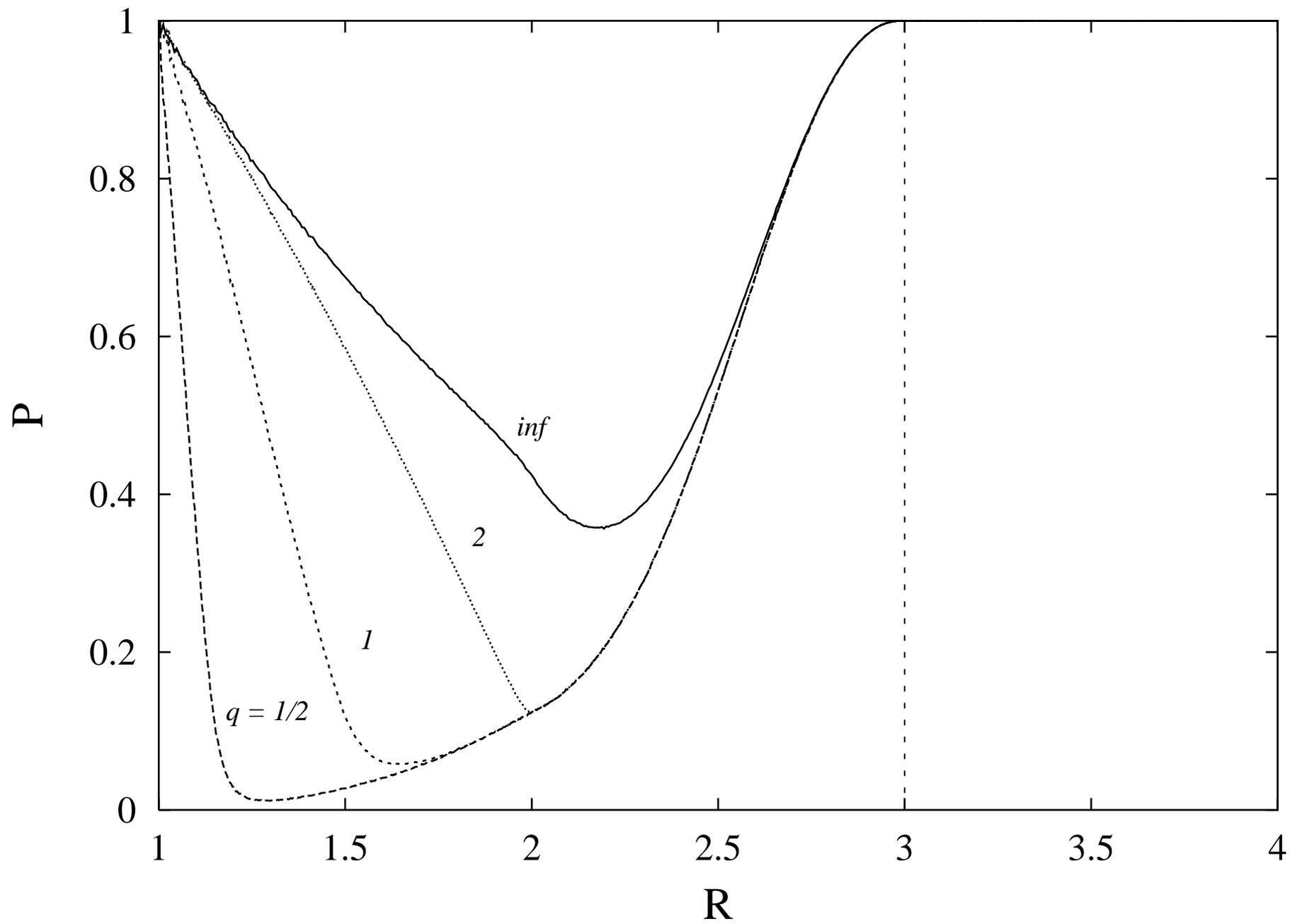
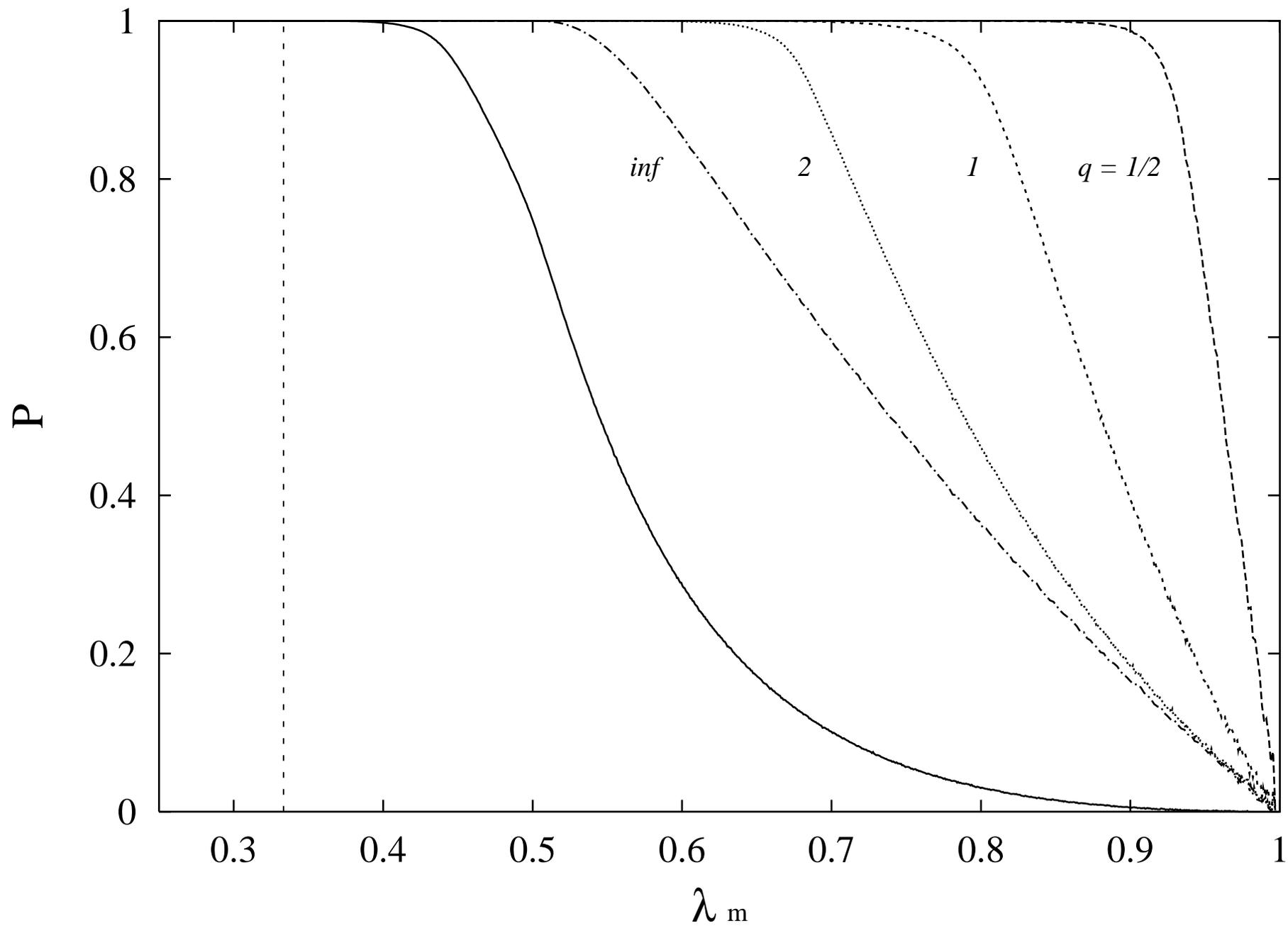
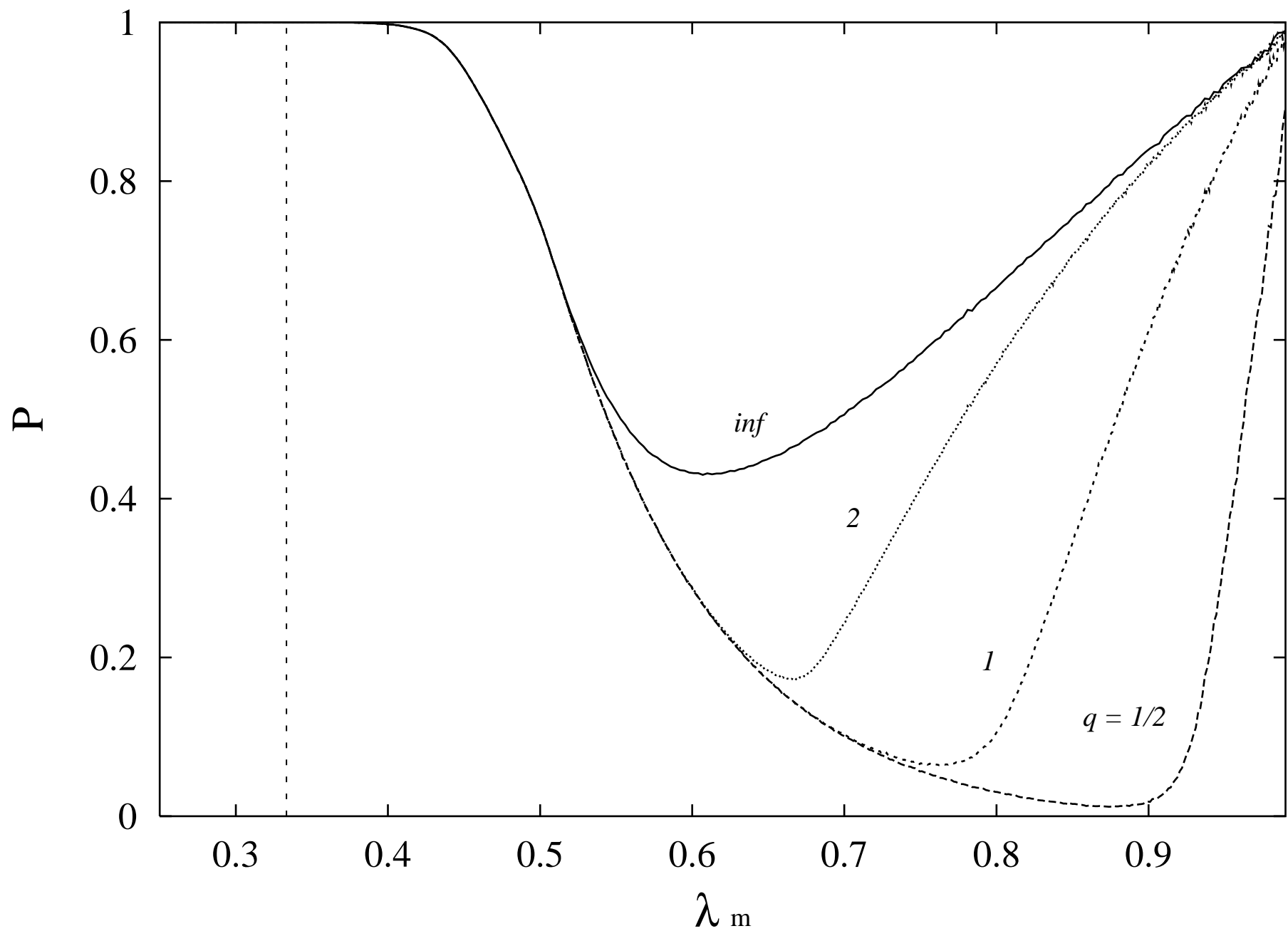
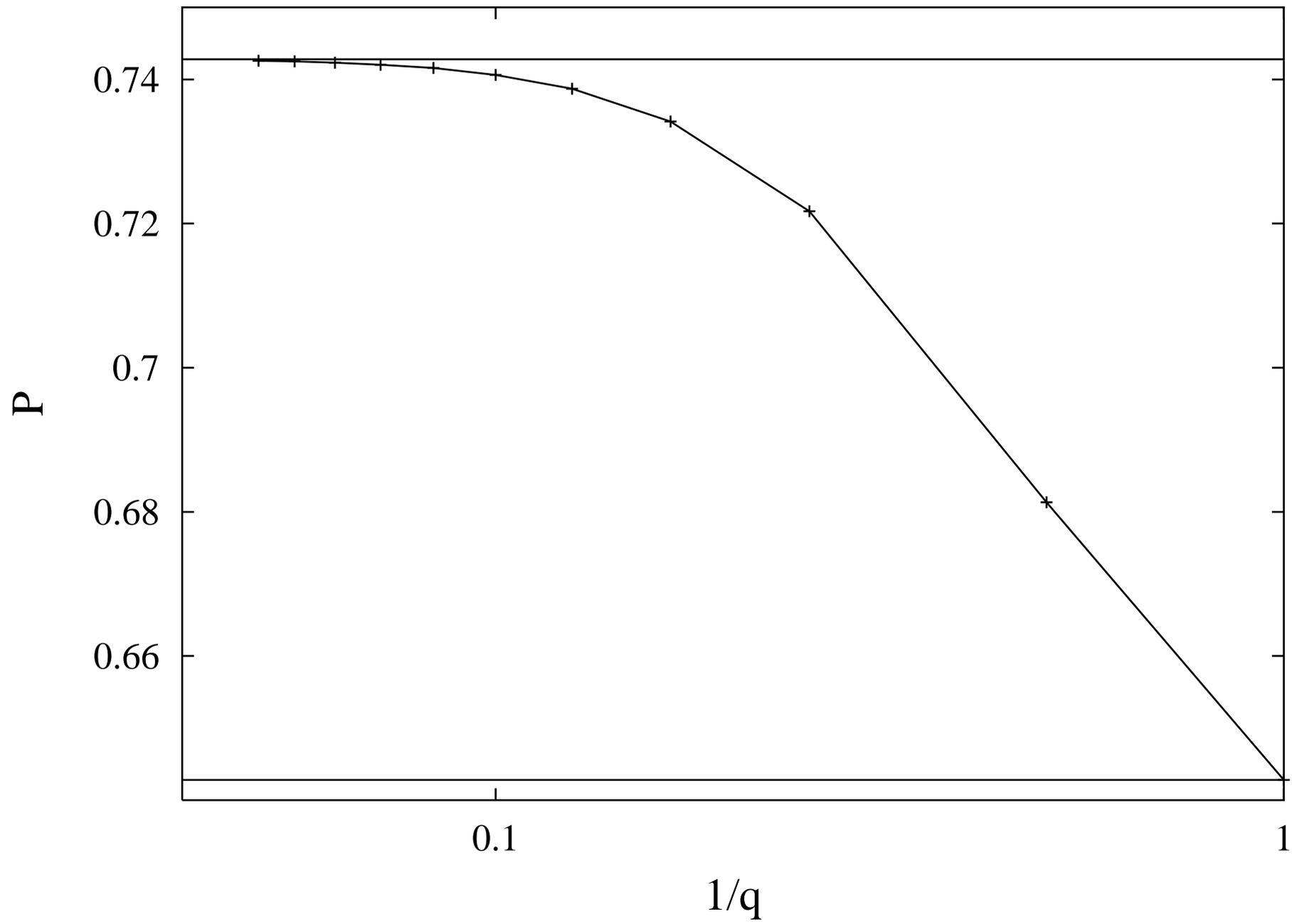
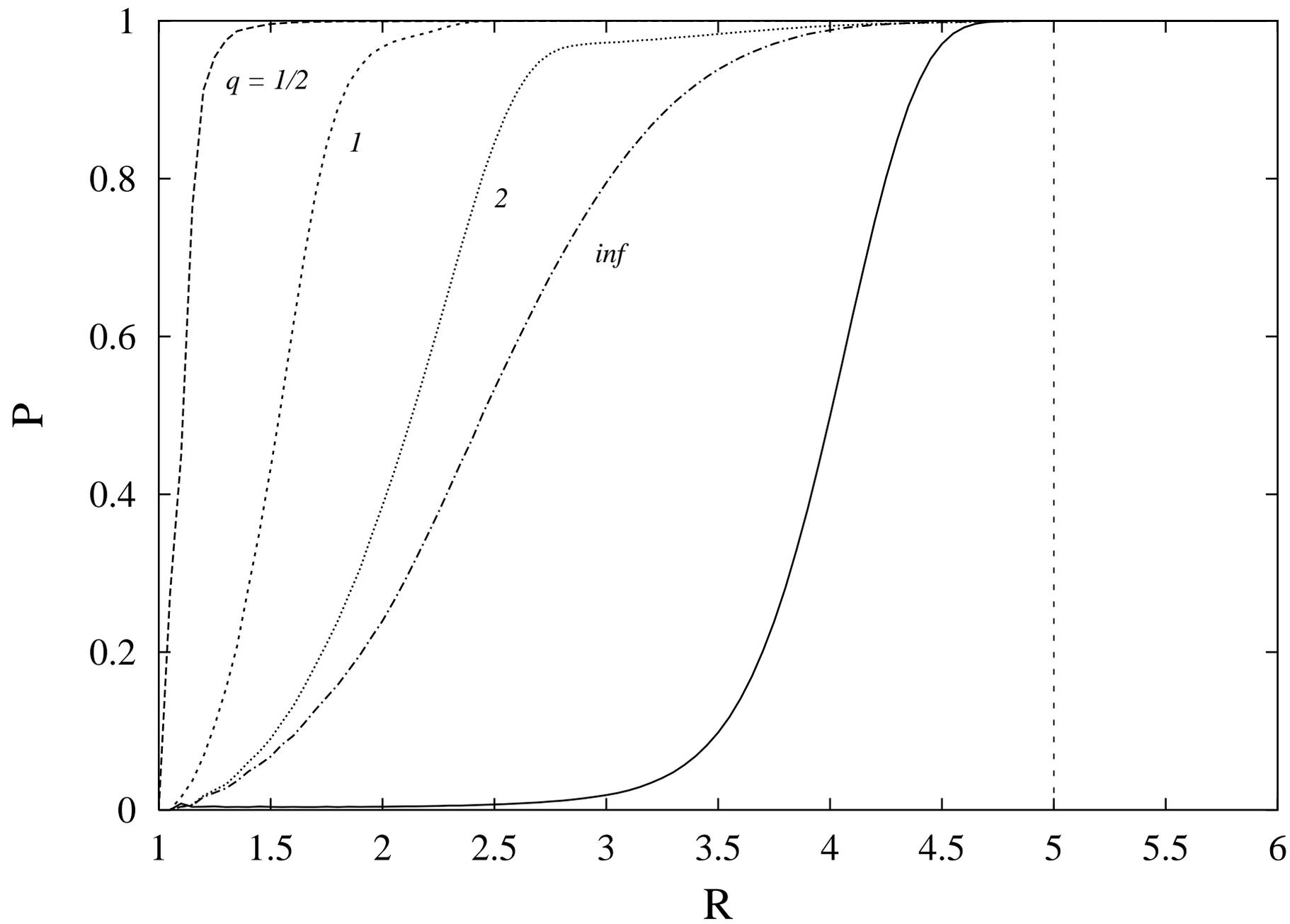
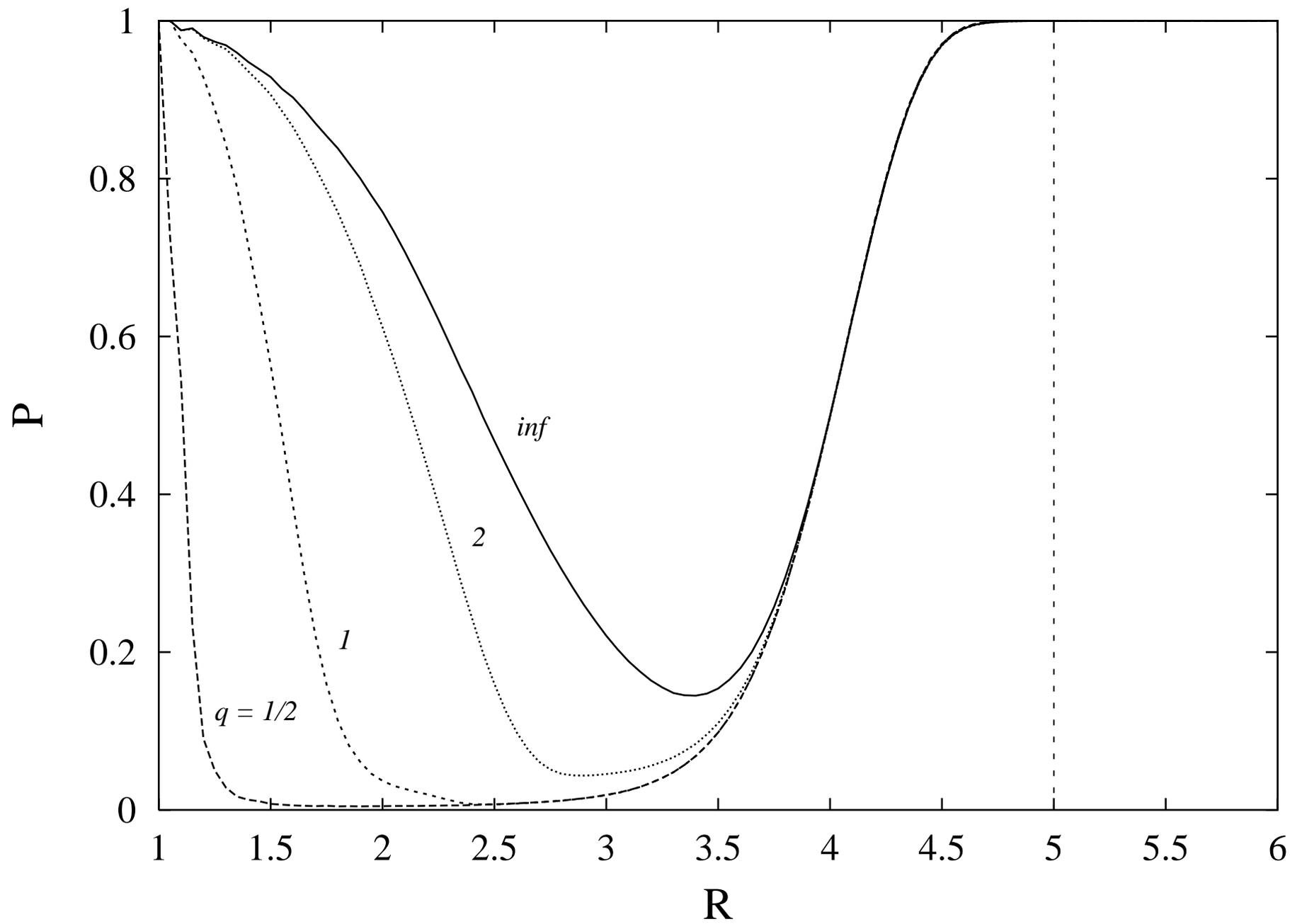
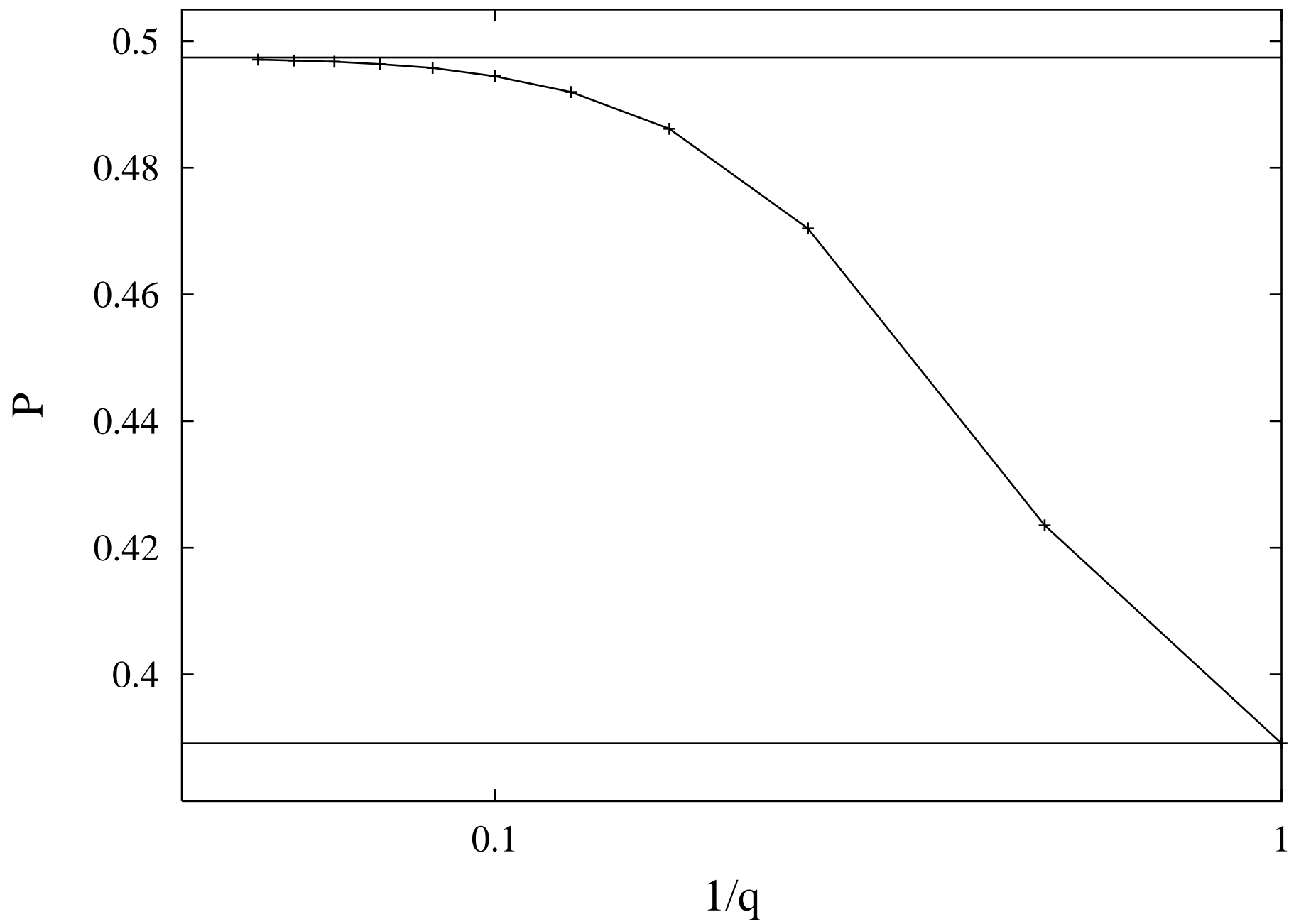
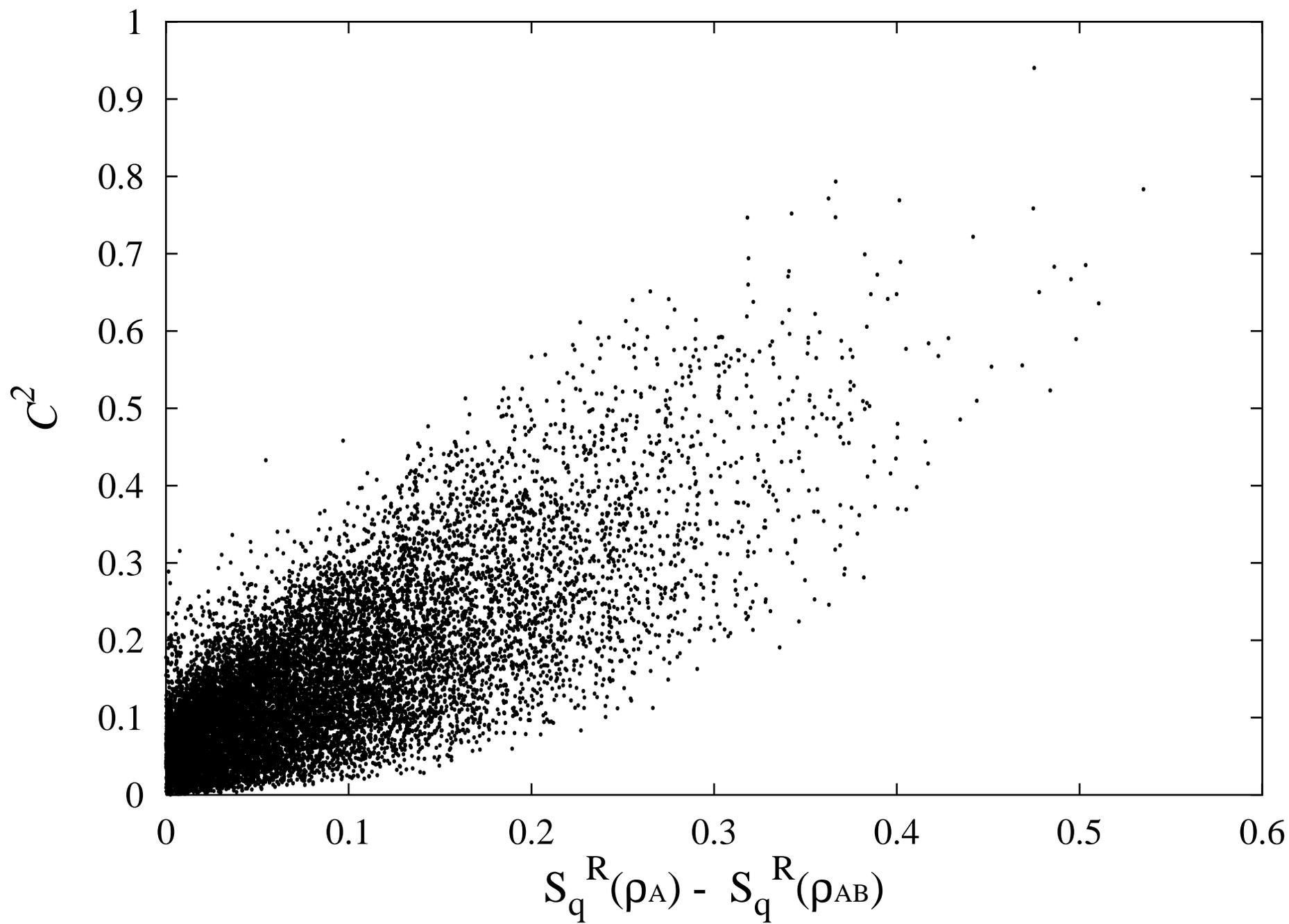
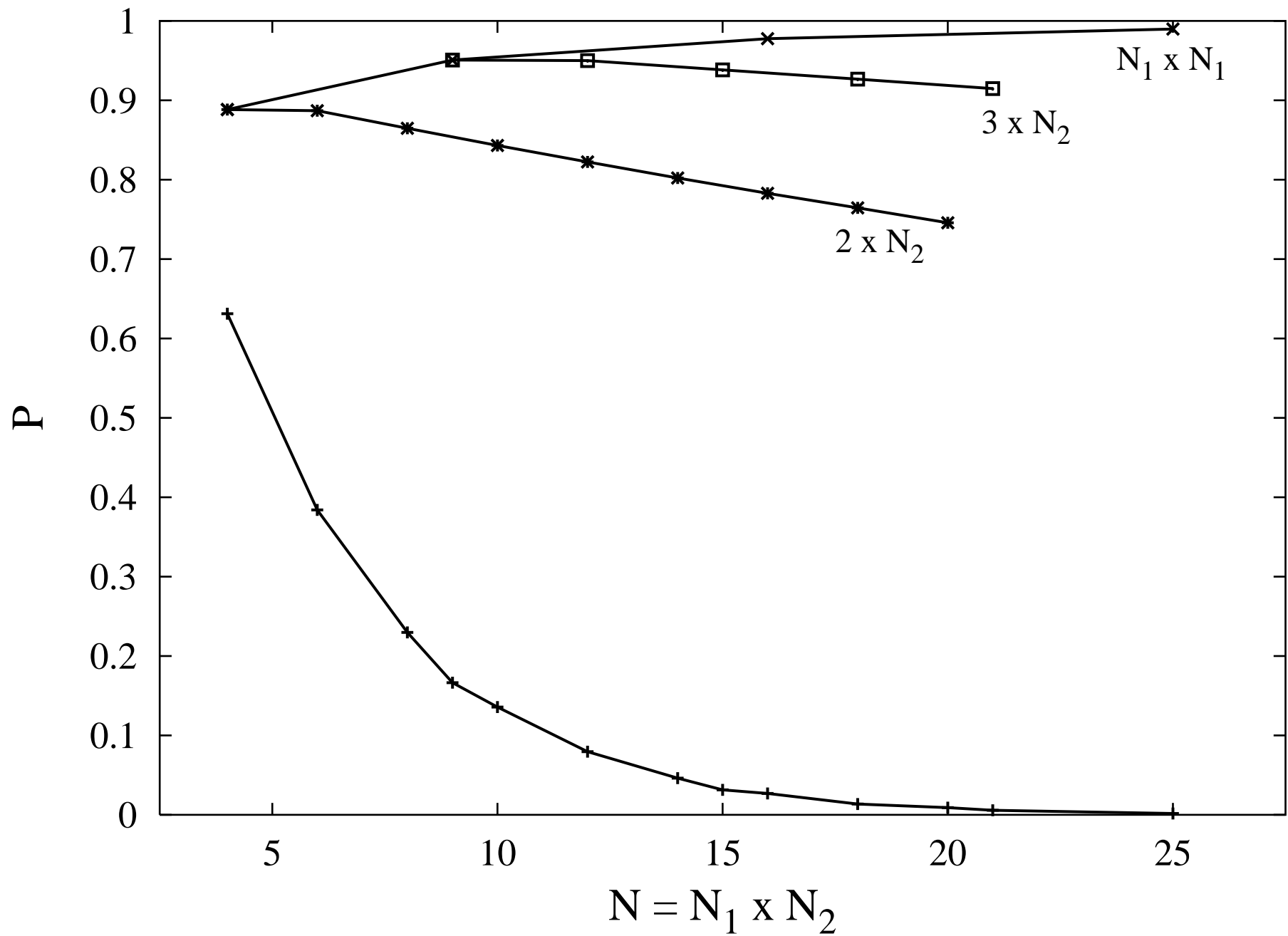
\end{document}